\documentclass[11pt]{article}

\usepackage{cite}
\usepackage{amssymb}

\def\bbox#1{\mbox{\boldmath$#1$}}

\begin{document}

\vspace{2cm}

\noindent
Preprint of the University of Technology, Dresden,\hfill
{\sf hep-ph/0111284}\\
and the Laboratoire Kastler Brossel \hfill
{\sf }\\[3ex]
\vspace{2cm}

\begin{center}
\begin{tabular}{c}
\hline
\rule[-5mm]{0mm}{15mm}
{\Large \sf Breit Hamiltonian and QED Effects for}\\[1ex]
{\Large \sf Spinless Particles}\\[2ex]
\hline
\end{tabular}
\end{center}

\vspace{1cm}

\begin{center}
U.~D.~Jentschura$^{a),b)}$, G.~Soff$^{a)}$ and P.~J.~Indelicato$^{b)}$
\end{center}

\vspace{0.2cm}

\begin{center}
$^{a)}${\it Institut f\"ur Theoretische Physik,}\\
{\it Technische Universit\"{a}t Dresden, 01062 Dresden, Germany} \\[1ex]
$^{b)}${\it Laboratoire Kastler Brossel, Case 74,
\'{E}cole Normale Sup\'{e}rieure et}\\
{\it Universit\'{e} P.~et M.~Curie, 4 place Jussieu, F-75252 Paris, France}
\end{center}

\vspace{1.3cm}

\begin{center}
\begin{minipage}{10.5cm}
{\underline{Abstract}} 
We describe a simplified derivation for the 
relativistic corrections of order $\alpha^4$ 
for a bound system consisting of two 
spinless particles. We devote special attention to pionium,
the bound system of two oppositely charged pions. The leading quantum 
electrodynamic (QED) correction to the 
energy levels is of the order of $\alpha^3$ and due to electronic
vacuum polarization. We analyze further corrections due to
the self-energy of the pions, and due to recoil effects,
and we give a complete result for the 
scalar-QED leading logarithmic corrections 
which are due to virtual loops involving only the scalar constituent
particles (the pions); these corrections are of order 
$\alpha^5 \,\ln \alpha$ for S states.
\end{minipage}
\end{center}

\vspace{1.3cm}

Exotic atoms and molecules (containing mesons, muons, and
other unusual particles)

\newpage

%
%
\section{Introduction}
\label{Introduction}

Exotic bound systems like pionium~\cite{AfEtAl1993,AfEtAl1994} 
(the bound system
of two oppositely charged pions) offer interesting possibilities
for studies of fundamental properties
of quantum mechanical bound states: the interplay
between strong-interaction corrections and quantum electrodynamic
corrections is of prime interest,
and the small length scales characteristic of the 
heavy particles make it possible to explore effects of the virtual 
excitations of the quantum fields in previously unexplored
kinematical regimes~\cite{KaJeIvSo1998,JeSoIvKa1997,KaIvJeSo1998}.
We do not wish to hide the fact that any potential high-precision
experiments in this area are faced with various 
experimental difficulties.
Our calculations address QED corrections to the spectrum 
of bound systems whose constituent particles are spinless;
relativistic corrections to the decay lifetime of pionium
have recently been discussed in~\cite{HeHeTrBa2001}
in the context of the DIRAC experiment at CERN.

Here, we report on results regarding the spectrum of a bound 
system consisting of two spinless particles.
We apply the simplified calculational scheme employed in
\cite{PaKa1995} for the relativistic and recoil corrections
to a bound systems of two ``non-Dirac'' particles to the case
of two interacting spinless particles (see Sec.~\ref{BreitEffects}). 
We then recall known results
on leading-order vacuum polarization corrections in Sec.~\ref{VacuumEffects}
and clarify the relative order-of-magnitude of the one- and two-loop
electronic vacuum polarization, the relativistic and recoil corrections
and the self-energy effects in pionium (also in Sec.~\ref{VacuumEffects}). 
We then provide
an estimate for the self-energy effect in Sec.~\ref{SelfEnergyEffects},
and we analyze the leading recoil correction of order
$\alpha^5$ (the Salpeter correction) which leads us to complete
results for the scalar-QED logarithmic corrections of order
$\alpha^5 \ln\alpha$.

%
%
\section{Breit Hamiltonian for Spinless Particles}
\label{BreitEffects}

We start from the Lagrangian for a 
charged spinless field coupled to 
the electromagnetic field [see equations~(6-50) -- 
(6-51b) of~\cite{ItZu1980}],
\begin{equation}
\label{Lagrangian}
{\mathcal L}(x) = 
\left[\left( \partial_\mu - {\mathrm i} e A_\mu \right) 
\phi^{*}(x) \right]\,
\left( \partial^\mu + {\mathrm i} e A^\mu \right) \phi(x) -
m^2 \phi^{*}(x) \, \phi(x) - 
\frac{1}{4} \, F_{\mu\nu}(x) \, F^{\mu\nu}(x)\,,
\end{equation}
where the field strength tensor $F_{\mu\nu}$ reads
$F_{\mu\nu}(x) = \partial_\mu A_\nu(x) - \partial_\nu A_\mu(x)$.
We use natural Gaussian units with $\hbar = c = \epsilon_0 = 1$.
The transition current for a free spinless particle 
($A^\mu = 0$) can be inferred from (\ref{Lagrangian}); it reads
in momentum space 
\begin{equation}
\label{currentjmu}
j^\mu(p',p) = \phi^{*}(p') \, (p'^\mu + p^\mu) \, \phi(p)\,.
\end{equation}
This current now has to be expressed in terms of nonrelativistic
wave functions. Specifically, the $j^0$-component has to reproduce
the normalization of the nonrelativistic (Schr\"{o}dinger)
wave function. By contrast, according to Eq.~(\ref{currentjmu})
the zero-component of the current reads $2 m\,\phi^{*} \phi$ 
in the nonrelativistic limit $p'^0 \to m\,, p^0 \to m$. 
The nonrelativistic wave functions
are normalized according to
\begin{equation}
\int {\mathrm d}^3 x \, \phi^{*}_{\mathrm S}(\bbox{x}) 
\phi_{\mathrm S}(\bbox{x}) = 1\,.
\end{equation}
It is therefore evident that we cannot simply associate the relativistic
wave function $\phi$ with $\phi_{\mathrm S}$; rather, we should 
define according to Eqs.~(13) -- (14) of~\cite{PaKa1995}
\begin{equation}
\phi(\bbox{p}) = 
\frac{\phi_{\mathrm S}(p^0,\bbox{p})}{\sqrt{2 p^0}}\,,
\end{equation}
where $p^0 = \sqrt{\bbox{p}^2 + m^2} \approx m$
is the energy of the free nonrelativistic
particle (in deriving low-energy
effective interactions, one always expands about {\em free}-particle
amplitudes; all interactions are treated as perturbations;
note the analogy to nonrelativistic QED -- NRQED -- 
for spinor particles~\cite{CaLe1986}).
The Klein--Gordon current, in the 
presence of external fields, reads in contrast to (\ref{currentjmu})
\begin{equation}
\label{currentjmuAmu}
j^\mu(p',p) = \phi^{*}(p') \, (p'^\mu + p^\mu - 2 \, e A^\mu) \, \phi(p)\,.
\end{equation}
The zero-component of this current can be interpreted
as a charge density, which is not necessarily positive definite.
Questions related to the normalization of the 
Klein--Gordon wave functions in this case are discussed in detail 
in~\cite{KlRa1975,BaLa1975,KlRa1975reply,FlSo1984}.

In terms of the Schr\"{o}dinger wave function,
the current is given as
\begin{eqnarray}
\label{j0}
j^0(\bbox{p}', \bbox{p}) &=&
\phi^{*}_{\mathrm S}(\bbox{p'}) \, 
\phi_{\mathrm S}(\bbox{p})\,, \\[2ex]
\label{ji}
j^i(\bbox{p}', \bbox{p}) &=& 
\phi^{*}_{\mathrm S}(\bbox{p'}) \, 
\frac{p^i + p'^i}{2\,m} \, 
\phi_{\mathrm S}(\bbox{p})\,,
\end{eqnarray}
where $m$ is the mass of the particle.
The atomic momenta $p^i$ and $p'^i$ in Eq.~(\ref{ji}) are
of order $Z\alpha$. As shown below, interactions
involving the spatial components $j^i$ of the
transition current give rise to relativistic
contributions of order $(Z\alpha)^4$ to the spectrum.
This is exactly the order of magnitude that is the subject of the
current investigation.
Therefore, although Eq.~(\ref{ji}) is only valid up to 
corrections of relative order $(Z\alpha)^2$, these 
can be neglected because the further corrections contribute
to the energy levels at the order of $(Z\alpha)^6$.
Specifically, we can expect corrections proportional to 
$(p^i \, \bbox{p}^2)$ to the current $j^i$ when a systematic expansion of the 
nonrelativistic current is performed; these terms are analogous 
to those obtained for relativistic corrections to the 
current of spinor particles which can be obtained via a
Foldy--Wouthuysen transformation~\cite{JePa1996,LaPa2001}.

In the following, the index S on the wave function will be 
dropped, and the nonrelativistic amplitudes describing the two interacting
particles (with electric charges $e_1$ and $e_2$) 
will be denoted as $\phi_1$ and $\phi_2$, respectively.
Following~\cite{PaKa1995},
the Breit Hamiltonian $U(\bbox{p}_1, \bbox{p}_2, \bbox{q})$
in momentum space 
is related to the invariant scattering amplitude $M$
and to the photon propagator $D_{\mu\nu}(q)$ in the following
way [see also equation (83,8) in~\cite{BeLiPi1991}]:
\begin{eqnarray}
M &=& e_1 \, e_2 \, j_1^\mu(\bbox{p}'_1, \bbox{p}_1) \, 
D_{\mu\nu}(\bbox{q}) \, 
j_2^\nu(\bbox{p}'_2, \bbox{p}_2) \nonumber\\[2ex]
&=& - \phi_1^{*}(\bbox{p}'_1) \, \phi_2^{*}(\bbox{p}'_2) \,
\left[ \frac{e_1\,e_2}{\bbox{q}^2} + U(\bbox{p}_1, \bbox{p}_2, \bbox{q})
\right]\,
\phi_1(\bbox{p}_1) \, \phi_2(\bbox{p}_2)
\end{eqnarray}
where $\bbox{q} = \bbox{p}'_2 - \bbox{p}_2 = - (\bbox{p}_1 - \bbox{p}'_1)$.
We employ a Coulomb-gauge photon propagator,
\begin{equation}
D_{00}(\bbox{q}) = - \frac{1}{\bbox{q}^2} \,, \quad
D_{ij}(\bbox{q}) = - \frac{1}{\bbox{q}^2 - \bbox{\omega}^2}\,
\left[ \delta^{ij} - \frac{q^i \, q^j}{\bbox{q}^2} \right]\,,
\end{equation}
where we can neglect the energy of the virtual photon for the 
derivation of next-to-leading order relativistic corrections,
\begin{equation}
D_{ij}(\bbox{q}) \approx - \frac{1}{\bbox{q}^2}\,
\left[ \delta^{ij} - \frac{q^i \, q^j}{\bbox{q}^2} \right]\,.
\end{equation}
The invariant scattering amplitude $M$ then reads
\begin{eqnarray}
\frac{M}{e_1\,e_2} &=& 
- \phi^{*}_1(\bbox{p}'_1) \, \phi^{*}_2(\bbox{p}'_2) \, \frac{1}{\bbox{q}^2}\,
\phi_1(\bbox{p}_1) \, \phi_2(\bbox{p}_2) \nonumber\\[1ex]
& & \;\;\; + \phi^{*}_1(\bbox{p}'_1) \, \phi^{*}_2(\bbox{p}'_2) \,
\left[ \frac{p^i_1 + p'^i_1}{2\, m_1} \,
\frac{p^j_2 + p'^j_2}{2\, m_2} \, \frac{1}{\bbox{q}^2} \,
\left[ \delta^{ij} - \frac{q^i \, q^j}{\bbox{q}^2} \right]\,
\right] \, \phi_1(\bbox{p}_1) \, \phi_2(\bbox{p}_2) \,.
\end{eqnarray}
We therefore identify
\begin{eqnarray}
\lefteqn{U(\bbox{p}_1, \bbox{p}_2, \bbox{q}) = 
- \frac{e_1\,e_2}{4 m_1 m_2} \, \frac{ (2 p_1^i - q^i) \,
(2 p_2^j + q^j)}{\bbox{q}^2} \,
\left[ \delta^{ij} - \frac{q^i \, q^j}{\bbox{q}^2} \right]}
\nonumber\\[2ex]
&=& - \frac{e_1\,e_2}{4 m_1 m_2} \, \left\{ 
\frac{(2 \bbox{p}_1 - \bbox{q}) \cdot (2 \bbox{p}_2 + \bbox{q})}{\bbox{q}^2} 
\right. \nonumber\\[1ex]
& & \;\;\; \left. - \frac{(2 \bbox{p}_1 \cdot \bbox{q} - \bbox{q}^2) \, 
(2 \bbox{p}_2 \cdot \bbox{q} + \bbox{q}^2)}{\bbox{q}^4} \right\} \,.
\nonumber\\[2ex]
&=&
- \frac{e_1\,e_2}{m_1 \, m_2} \, 
\left[ \frac{\bbox{p}_1 \cdot \bbox{p}_2}{\bbox{q}^2} 
- \frac{(\bbox{p}_1 \cdot \bbox{q}) \,
(\bbox{p}_2 \cdot \bbox{q})}{\bbox{q}^4} \right]\,.
\end{eqnarray}
We now transform to the center-of-mass frame in which 
$\bbox{p}_1 = -\bbox{p}_2 = \bbox{p}$, so that the expression
for $U(\bbox{p}_1, \bbox{p}_2, \bbox{q})$ becomes even simpler,
\begin{eqnarray}
\label{penultimate}
U(\bbox{p}, -\bbox{p}, \bbox{q}) =
\frac{e_1\,e_2}{m_1 \, m_2} \, 
\left[ \frac{\bbox{p}^2}{\bbox{q}^2} 
- \frac{(\bbox{p} \cdot \bbox{q})^2}{\bbox{q}^4} \right]\,.
\end{eqnarray}
The formula (83,13) of~\cite{BeLiPi1991}
can now be employed in evaluating the Fourier transform,
\begin{equation}
\int \frac{{\mathrm d}^3 q}{(2 \pi)^3} \,
\exp\left({\mathrm i} \, \bbox{q} \cdot \bbox{r} \right) \,
\frac{4 \pi (\bbox{a} \cdot \bbox{q}) \, (\bbox{b} \cdot \bbox{q})}{\bbox{q}^4}
= \frac{1}{2 r}\,\left[ \bbox{a} \cdot \bbox{b} -
\frac{(\bbox{a} \cdot \bbox{r}) (\bbox{b} \cdot \bbox{r})}{r^2} \right]\,.
\end{equation}
The Breit Hamiltonian, which we would like to 
denote by $H_{\mathrm B}$, is obtained by adding to the
Fourier transform of (\ref{penultimate})
the relativistic correction to the kinetic energy.
Denoting with $\bbox{\hat{p}} = - {\rm i} \,
{\partial}/{\partial \bbox{x}}$ the momentum operator in the
coordinate-space representation, we obtain 
\begin{eqnarray}
\label{hbreit}
H_{\mathrm B}(\bbox{r}, \bbox{\hat{p}}) =
- \frac{\bbox{\hat{p}}^4}{8\,m_1^3} - \frac{\bbox{\hat{p}}^4}{8\,m_2^3} 
+ \frac{e_1 \, e_2}{8 \pi r} \, \frac{\bbox{\hat{p}}^2}{m_1 \, m_2} 
+ \frac{e_1 \, e_2}{8 \pi r^3} \, \frac{\bbox{r} \cdot
(\bbox{r} \cdot \bbox{\hat{p}}) \,\, \bbox{\hat{p}}}{m_1 m_2}\,.
\end{eqnarray}
In the order of $(Z\alpha)^4$,
there is no contribution due to virtual annihilation 
for spinless particles; corrections of this type would enter only for 
positronium and dimuonium~\cite{KaIvJeSo1998} because they 
are caused by the spin-dependent part of the transition
current [see Eqs.~(83,20) and (82,22) of~\cite{BeLiPi1991}],
which is absent for spinless particles. For S states, virtual
annihilation is altogether prohibited by angular momentum conservation.

The matrix elements of the 
Breit Hamiltonian (\ref{hbreit}) for spinless particles can be evaluated
on nonrelativistic bound states via computational techniques
outlined in Sec.~A3 of Ch.~1 of~\cite{BeSa1957}.
For $m_1 = m_2 = m$ and $e_1\,e_2 = - 4 \pi Z \alpha$,
we obtain 
\begin{equation}
\label{NLresult}
E_{\mathrm{B}} = - \frac{(Z\alpha)^2 \, m}{4 n^2} 
- \frac{(Z\alpha)^4 \, m}{2 \, n^3} \,
\left[ \frac{1}{2 l + 1} - \frac{1}{4} \, \delta_{l0} - 
\frac{11}{32\,n} \right]
\end{equation}
as the Breit energy for 
the energy levels of the bound system of two 
spinless particles, including relativistic 
corrections of order $(Z \alpha)^4$.
Here, we keep $Z$ as a parameter which denotes the nuclear charge number
in a bound system. Of course,
for two particles each carrying an elementary charge,
$Z$ has to be set to unity. The fine-structure constant is 
denoted by $\alpha$. The result (\ref{NLresult}) agrees
with previous calculations~\cite{BaGl1955,BrItZJ1969,Na1972,Ow1994,%
HaOw1994,PaKa1995}, notably with Eq.~(38) of~\cite{Na1972}. 

It is instructive to compare the result (\ref{NLresult}) with the 
known result for a single-particle system of mass $m/2$ satisfying
the Klein--Gordon equation, bound to a nucleus with 
charge $Z e$. According to Eq.~(2-86) of~\cite{ItZu1980}, 
we obtain the ``Klein--Gordon energy'' (KG) 
\begin{equation}
\label{KGresult}
E_{\mathrm{KG}} = - \frac{(Z\alpha)^2 \, m}{4 n^2}
- \frac{(Z\alpha)^4 \, m}{2 \, n^3} \,
\left[ \frac{1}{2 l + 1} - \frac{3}{8\,n} \right]\,.
\end{equation}
The two results (\ref{NLresult}) and
(\ref{KGresult}) are manifestly different in the order of
$(Z\alpha)^4$.

>From (\ref{hbreit}) we conclude that the zitterbewegung term is absent
for spinless particles. However, this statement is in need of further
explanation because a considerable variety of physical
interpretations exists in the literature
with regard to the zitterbewegung term. 
We briefly expand:
The Dirac $\alpha$-matrices fulfills $\alpha = {\mathrm i} [H_{\mathrm D},x]$ 
($H_{\rm D}$ is the Dirac hamiltonian)
as the relativistic generalization of the velocity operator. 
By contrast, in the 
nonrelativistic formalism, we have the analogous 
relation $p/m  = {\mathrm i} [H_{\mathrm S} ,x]$ where $H_{\mathrm S}$
is the Schr\"{o}dinger Hamiltonian.
Since the $\bbox{\alpha}$--matrices have eigenvalues $\pm 1$, the 
magnitude of the velocity of the electron -- at face value --
is equal to the velocity
of light at any given instant. On p.~106 of~\cite{He1950}, it is argued that
``the explanation for this fact is that the electron carries out a fast 
irregular motion (``zitterbewegung'') -- which is responsible for the 
spin -- whereas the mean velocity is given by the momentum
$\bbox{p}/m$''.  Note that the introduction of the Dirac matrix 
formalism is necessitated by the need to describe the internal degrees
of freedom of the particle -- the spin.
On p.~71 of~\cite{ItZu1980}, it is shown that
the zitterbewegung term can be traced to the positional
fluctuations $\langle \delta \bbox{r}^2 \rangle \sim 1/m^2$ of the electron,
and a connection is drawn to the Darwin term which results naturally 
in the context of the Foldy--Wouthuysen transformed Dirac hamiltonian.
On pp.~117--118 of~\cite{Sa1967Adv} and 
p.~62 of~\cite{ItZu1980}, it is argued that the momentum 
$\bbox{p}$ of a Dirac wave packet can be associated in a natural
way with the group velocity, but that in addition to the group
velocity term, there exist highly oscillatory terms which represent
the zitterbewegung.  
Similarly, on pp.~139--140 of~\cite{Sa1967Adv}, it is shown that
the zitterbewegung term can also be interpreted as arising from
the interaction of the atomic electron with virtual electron-positron
pairs created in the Coulomb 
field of the nucleus. This virtual electron-positron
pair-creation is subject to the uncertainty principle and can
occur only for time intervals of the order of
$\Delta t \sim \hbar/(2 m c^2)$ (where we temporarily restore
the factor $\hbar$). At the time the original atomic electron
fills up the vacated negative-energy state (the bound-electron wave-function
has negative-energy components), the escalated electron (which forms
part of the virtual pair) is at most 
a distance $c \Delta t \sim 1/m$ away from the original electron.
This distance is precisely of the order of magnitude of the 
fluctuations of the electron coordinate and consistent
with the discussion on p.~71 of~\cite{ItZu1980}.
All these interpretations elucidate different
aspects of the same problem.

In the context of the Breit hamiltonian,
we would like to adhere to the definition
that the zitterbewegung term is the term of order $(Z\alpha)^4$ in 
the Breit Hamiltonian generated by a contribution which is manifestly
proportional of $\delta(\bbox{r})$ in coordinate space 
(or a constant in momentum space). Such a term is absent in 
the result (\ref{hbreit}).  
For spin-$1/2$ particles, such a term is generated by the multiplication
of the photon propagator (proportional to $1/\bbox{q}^2$) with 
the zero-component of the transition current  which is given
for a spin-$1/2$ particle as [see equation (4) of~\cite{PaKa1995}]
\begin{equation}
\bar{u}' \gamma^0 \bar{u} =
w^{*} \, \left( 1 - \frac{\bbox{q}^2}{8\,m^2} +
\frac{{\mathrm i}\, \bbox{\sigma} \cdot \bbox{p}' \times \bbox{p}}
  {4\,m^2} \right) \, w\,.
\end{equation}
Here, $u$ is the bispinor amplitude for the bound particle,
and $w$ is the bound-state Schr\"{o}dinger wave function related by
\begin{equation}
u = \left( \begin{array}{c} 
\left(1 - 
\frac{\displaystyle \bbox{p}^2}{\displaystyle 8\,m^2}\right) w \\[3ex]
\frac{\displaystyle \bbox{\sigma}\cdot\bbox{p}}{\displaystyle 2 m} w 
\end{array} \right)
\end{equation}
according to equation (3) of~\cite{PaKa1995}.
One might wonder why a term proportional to $\delta_{l0}$,
apparently generated by a $\delta$-function in coordinate space,      
prevails in the Breit energy (\ref{NLresult}).
This term arises naturally when evaluating a matrix element
of the structure 
$\langle \phi_{\mathrm S} |
(\bbox{r} \cdot (\bbox{r} \cdot \bbox{\hat{p}}) \,
\bbox{\hat{p}})/r^3 | \phi_{\mathrm S} \rangle$
(last term of equation~(\ref{hbreit}))
on the nonrelativistic wave function $\phi_{\mathrm S}$ and
should {\em not} be associated with the zitterbewegung.

Vacuum polarization corrections and self-energy effects, as well
as corrections due to the strong interaction, are not included in 
(\ref{NLresult}). These corrections will be discussed in the two 
following sections.

%
%
\section{Vacuum Polarization Effects}
\label{VacuumEffects}

As pointed out by various
authors (e.g.~\cite{Pu1957,
BoRi1982,EiSo2000plb,EiSo2000prd,LaBu1998,JeSoIvKa1998,KaJeIvSo1998EPJ}), 
the electronic vacuum polarization enters already at the order of $\alpha^3$ 
[more precisely, $\alpha\,(Z\alpha)^2$]
in bound systems with spinless particles, because the spinless 
particles are much heavier than the electron, which means that 
the Bohr radius of the bound system is roughly
of the same order of magnitude
as the Compton wavelength of the electron.
The Compton wavelength of the electron, however, is the fundamental 
length scale at which the charge of any bound particle is screened
by the electronic vacuum polarization. 

The vacuum polarization (VP) correction to energy levels 
has been evaluated~\cite{Pu1957,EiSo2000plb,EiSo2000prd,
JeSoIvKa1998,KaIvJeSo1998,LaBu1998}
with nonrelativistic wave functions. We recall
that the leading-order VP correction (due to the Uehling potential)
can be expressed as
\begin{equation}
\label{defVP}
\Delta E = \langle \psi | V_U | \psi \rangle =
\frac{\alpha}{\pi}\,C_E\,E_\psi\,,
\end{equation}
where
\begin{equation}
\label{Schroedinger}
E_\psi = -\frac{(Z\,\alpha)^2\,m}{4\,n^2}
\end{equation}
is the Schr\"odinger binding energy for a two-body system with two 
particles each of mass $m$ [first term on the right-hand side
of (\ref{NLresult})]. For the $C_E$ coefficients, we recall
the following known results~\cite{EiSo2000plb,EiSo2000prd,KaIvJeSo1998},
\begin{equation}
\label{resVP}
C_E({\mathrm{1S}}) = 0.22 \,, \quad
C_E({\mathrm{2S}}) = 0.10\,.
\end{equation}
As an alternative to the
nonrelativistic treatment,
the Uehling correction could also be evaluated with 
relativistic bound-state 
Klein--Gordon wave functions (although these do not describe
the two-body system accurately, as shown in Sec.~\ref{BreitEffects}).
The difference could be
interpreted as a rough estimate of further relativistic 
corrections not taken into account in the nonrelativistic treatment
of the vacuum polarization.
The result obtained by numerically solving the Klein--Gordon equation
and numerically evaluating the Uehling correction is shown  
in the sixth row of Tab.~\ref{table1}. This relativistic result
is in very good agreement with Eq.~(\ref{resVP}). 
As pointed out in~\cite{GaLyRuGa2001}, the strong interaction correction
is also an $\alpha^3$ effect (like the vacuum polarization)
and enters at a relative order of $\alpha$, i.e.~on the level of 
about $1\,\%$ in pionium.

We recall here also the known results for the vacuum polarization correction
to the charge density at the origin~\cite{JeSoIvKa1998,LaBu1998},
\begin{equation}
\label{respi}
\left[\frac{\Delta|\psi_{\rm 1S}(0)|^2}
  {|\psi_{\rm 1S}(0)|^2}\right]_{\pi^{+}\pi^{-}}
=  \frac{\alpha}{\pi}\,1.36 \;\;\; \mbox{and} \;\;\;
\left[\frac{\Delta|\psi_{\rm 2S}(0)|^2}
  {|\psi_{\rm 2S}(0)|^2}\right]_{\pi^{+}\pi^{-}}
=  \frac{\alpha}{\pi}\,1.14\,.
\end{equation}
Two-loop vacuum polarization effects enter at a 
relative order $\alpha^2$ in pionium and are therefore 
of the same order of magnitude as the relativistic 
corrections mediated by the Breit interaction (discussed in
Sec.~\ref{BreitEffects} and termed ``two-body correction'' in
Tab.~\ref{table1}). The self-energy correction which is 
discussed in the following section is even smaller, but of considerable
theoretical interest. 

%
%
\section{Effects due to Scalar QED}
\label{SelfEnergyEffects}

As shown in~\cite{Pa1998}, the leading logarithmic correction
to the self energy can be obtained, in nonrelativistic approximation,
from second-order perturbation theory based on
nonrelativistic quantum electrodynamics~\cite{CaLe1986}
(see also~\cite{BB1984}). We will investigate here, in a
systematic way, the leading logarithms generated for 
S states by self-energy and relativistic-recoil effects 
(the so-called Salpeter correction), and show that these
are spin-independent.

The quantized electromagnetic field 
is [see Eq.~(5) of~\cite{Pa1998}],
\begin{equation}
\bbox{A}(\bbox{r}) = \sum_{\lambda=1,2} 
\int \frac{{\mathrm d}^3 k}{\sqrt{(2\pi)^3 \, 2 k}}\,
\bbox{\epsilon}_{\lambda}(\bbox{k}) \,
\left[ a^{+}_{\bbox{k},\lambda} \exp(-{\mathrm i} \bbox{k}\cdot\bbox{r}) +
a_{\bbox{k},\lambda} \exp({\mathrm i} \bbox{k}\cdot\bbox{r})\right]\,,
\end{equation}
and the nonrelativistic interaction Hamiltonian for an atomic
system with two spinless particles (charges $e_1$ and $e_2$ and
masses $m_1$ and $m_2$) reads
\begin{equation}
\label{HI}
H_{\mathrm I} = - \frac{e_1}{m_1} \, \bbox{p_1}\cdot \bbox{A}(\bbox{r}_1) + 
\frac{e_1^2}{2 m_1} \bbox{A}(\bbox{r}_1)^2 
- \frac{e_2}{m_2} \, \bbox{p_2}\cdot \bbox{A}(\bbox{r}_2) +
\frac{e_2^2}{2 m_2} \bbox{A}(\bbox{r}_2)^2 \,.
\end{equation}
For two spin-$1/2$ particles, the terms
\begin{equation}
\label{HIS}
- \frac{e_1}{m_1} \, \bbox{\sigma}_1 \cdot \bbox{B}(\bbox{r}_1)
- \frac{e_2}{m_2} \, \bbox{\sigma}_2 \cdot \bbox{B}(\bbox{r}_2)
\end{equation}
have to be added to $H_I$ [see Eq.~(7) of~\cite{Pa1998}]. 
We will carry out the calculations for the general case of one
particle of charge $e_1 = e$ and the other having a charge $e_2 = - Z e$
(we follow the convention of~\cite{ItZu1980}
that for hydrogen, $e$ is the physical charge of the electron,
i.e.~$e = - |e|$).
The unperturbed Hamiltonian of the system of the
two particles and the electromagnetic
field reads [see e.g.~Eq.~(6) of~\cite{Pa1998}],
\begin{equation}
\label{H0}
H_0 = \frac{\bbox{p}_1^2}{2 m_1} +
\frac{\bbox{p}_2^2}{2 m_2} - \frac{Z\alpha}{r} +
\sum_{\lambda=1,2} \int {\mathrm d}^3 k \, k \,
a^{+}_{k,\lambda} \, a_{k,\lambda} \,.
\end{equation}
where $\bbox{r} = \bbox{r}_1 - \bbox{r}_2$.
The eigenstates of the ``atomic part'' $H_0^A$ of this Hamiltonian 
in the center-of-mass system $\bbox{p}_1 + \bbox{p}_2 = 0$ are the 
nonrelativistic Schr\"{o}dinger--Coulomb wave functions for a reduced
mass $m_{\rm r} = m_1 \, m_2 / (m_1 + m_2)$
[here,  the ``atomic part'' $H_0^A$ excludes the photon field,
i.e.~the last term of~(\ref{H0})]. We denote
$\bbox{p} \equiv \bbox{p}_1 = -\bbox{p}_2$.

Given that the first-order perturbation
$\langle \phi_{\mathrm S} | H_{\mathrm I} | \phi_{\mathrm S}\rangle$
vanishes, the second-order perturbation yields the dominant
nonvanshing perturbation. When evaluated on an atomic state, it 
is given by
\begin{equation}
\label{2ndorder}
\delta E_{\mathrm{SE}} = \langle \phi_{\mathrm S} | H_{\mathrm I}
\frac{1}{H_0 - E_{\mathrm S}} H_{\mathrm I} | \phi_{\mathrm S} \rangle\,.
\end{equation}
The interaction Hamiltonian (\ref{HI}) gives rise to 
QED corrections that involve both
particles (in the current context, these are
recoil corrections involving the product $e_1 \, e_2$), 
and also to terms which involve only a single particle and
are proportional to $e_1^2$ or $e_2^2$. The latter effects 
correspond to the self-energies of the two particles.

The low-energy part of the self-energy in leading order~\cite{Pa1993} 
can be inferred directly from
(\ref{2ndorder}), and it can be seen
that the spin-dependent parts from (\ref{HIS}) vanish
in the leading order in the $(Z\alpha)$-expansion~\cite{Pa1998}:
\begin{equation}
\label{NREL}
E_{\mathrm L} = -\frac{e_1^2}{6 \, \pi^2} \,
\int_0^\epsilon {\mathrm d}k \, k \,
  \left< \phi \left| \frac{\bbox{p}}{m_1} \,\,\, 
    \frac{1}{H_0^A - (E_{\mathrm S} - k)} \,\,\,
     \frac{\bbox{p}}{m_1} \right| \phi \right> +
(e_1 \leftrightarrow e_2, m_1 \leftrightarrow m_2)\,.
\end{equation}
where 
\begin{equation}
E_{\mathrm S} = -\frac{(Z\alpha)^2 m_{\mathrm r}}{2 n^2}
\end{equation}
is the Schr\"{o}dinger energy ($m_{\mathrm r}$ in the reduced mass
of the atomic system under investigation). 
Starting from the {\em spin-independent} expression (\ref{NREL}),
it is now relatively straightforward to show
that the leading ``self-energy logarithm'' for S states is given by
\begin{equation}
\label{SEestimate}
\delta E_{\mathrm{SE}} \approx 
\frac{4 \, \ln(Z\alpha)^{-2}}{3\,\pi\,n^3} \, 
\delta_{l0}\, \left[\alpha \, (Z \alpha)^4 \, \frac{m^3_{\mathrm r}}{m_1^2} 
+ Z \, (Z \alpha)^5 \, \frac{m^3_{\mathrm r}}{m_2^2}\right] \,.
\end{equation}
This result is by consequence spin-independent.
The derivation is simplified when using the $\epsilon$-method developed
and used in various bound-state calculations~\cite{Pa1993,JePa1996,JeSoMo1997}.
The two terms in square brackets in (\ref{SEestimate}) 
correspond to the two self-energies of the two constituent particles
with charges $e_1 = e$ and $e_2 = -Z e$ and masses $m_1$ and $m_2$,
respectively. It has been pointed out~\cite{SaYe1990}
that in contrast to the self-energy corrections,
the vacuum polarization corrections given in Eq.~(\ref{respi}) 
must not be double-counted. The ``double-counting'' of 
self-energy corrections (and lack of it in the vacuum-polarization
case) finds a natural explanation in our formalism: whereas the 
vacuum-polarization correction mainly leads to a modification of the
$1/r$-type Coulomb attraction in (\ref{H0}) within a nonrelativistic
effective theory, the structure of the interaction
Hamiltonian (\ref{HI}) implies the existence of the 
{\em two} self-energies of the {\em two} constituent particles
of the atomic system.

It might be instructive to point out that
the formula (\ref{SEestimate}) is consistent 
with Welton's argument for estimating the self-energy effect 
on a bound particle
which is based on analyzing the influence of the fluctuating
electromagnetic field [a detailed discussion is given on pp.~80--82 
of~\cite{ItZu1980}]. 
For a system with two particles of equal mass
$m_1 = m_2 = m$, we have 
$m_{\mathrm r} = m/2$.

The leading-order recoil correction (Salpeter correction)
can also be inferred from the
interaction Hamiltonian (\ref{HI}) via second-order perturbation
theory, by ``picking up'' terms that involve
products $e_1 \, e_2$.
It has been shown in~\cite{Pa1998} that the leading {\em logarithm}
(for S states) of the Salpeter correction 
is spin-independent (just like the leading logarithm of the 
self-energy correction). The Salpeter correction
is usually referred to as a relativistic recoil (RR) correction.
By following~\cite{Pa1998}, we obtain for the leading logarithm of this
effect
\begin{equation}
\label{RRestimate}
\delta E_{\mathrm{RR}} \approx 
\frac{2\,(Z \alpha)^5}{3\,\pi\,n^3} \, \delta_{l0} \,
\ln\left(\frac{1}{Z\alpha}\right) \, \frac{m^3_{\mathrm r}}{m_1 m_2} \,.
\end{equation}
This correction involves only 
the products $e_1 \, e_2 = - 4 \pi Z\alpha$ and can therefore
be written as a function of $Z\alpha$ alone.
 
For pionium, we have $Z=1$, $m_1=m_2=m=m_\pi$, $m_{\mathrm r} = m_\pi/2$.
The leading logarithmic correction from scalar QED for 
pionium in the order of $\alpha^5 \ln\alpha$ is obtained
by adding the corrections (\ref{SEestimate}) and (\ref{RRestimate}),
\begin{equation}
\label{reslog}
\delta E_{\mathrm{log}} = \delta E_{\mathrm{SE}} + 
  \delta E_{\mathrm{RR}} 
  = \frac{3}{4} \, 
    \frac{\alpha^5}{\pi\,n^3} \, \ln\left(\frac{1}{\alpha}\right) \, m_\pi\,.
\end{equation}
The non-logarithmic term of order $\alpha^5$ is spin-dependent,
and its evaluation requires a relativistic treatment of 
the self-energy effect
of a bound spinless particle; such a calculation
would be of considerable theoretical
interest, but the size of the effect for pionium, which is roughly 
two orders of $\alpha$ smaller than the leading vacuum polarization 
correction, precludes experimental verification in the near future.
However, we would like to point out here that a fully relativistic
treatment of this problem, including a detailed discussion of the 
renormalization of the self energy of the spinless particle, has 
not yet been accomplished. Scalar QED is a renormalizable 
theory~\cite{ItZu1980}.

The dominance of vacuum polarization over self-energy effects
in pionium is expressed, in particular, by the fact that even
{\em two-loop} vacuum polarization of order $\alpha^4$ has
a stronger effect on the spectrum of pionium than the leading
logarithm from Eq.~(\ref{reslog}), and that the strong-interaction
correction of order $\alpha^3$~\cite{GaLyRuGa2001} has to be well
understood before any experimental verification of~(\ref{reslog})
appears feasible.
Finally, we remark that for a manifestly non-elementary particle
like the pion which has a finite charge radius, form-factor 
corrections have to be taken into account. 

\begin{table}[htbp]
\begin{center}
\begin{minipage}{12cm}
\begin{center}
\caption{QED contributions to the $1{\mathrm S}$ level of 
pionium in eV. For a further discussion of the corrections
see the text.}
\vspace*{0.3cm}
\label{table1}
\begin{tabular}{lr}
\hline
\hline
Contribution & Energy (eV) \\
\hline
One--body Klein--Gordon [Eq.~(\ref{KGresult})]
                                  & -1858.19895 \\[1ex]
Higher Order Klein--Gordon 
[Exact -- Eq.~(\ref{KGresult})] 
                                  & -0.00001 \\[1ex]
Form factor correction to Klein--Gordon (0.61 fm) 
                                  & 0.01308 \\[1ex]
Two-body correction (Breit)
[Eq.~(\ref{NLresult}) -- 
Eq. (\ref{KGresult})]             & 0.04329 \\[1ex]
Uehling (with a relativistic wave function) 
                                  & -0.94228 \\[1ex]
Vac. Pol. (Wichmann-Kroll)        & 0.00001 \\[1ex]
Vac. Pol. (K{\"a}ll{\'e}n-Sabry)  & -0.00729 \\[1ex]
Vac. Pol. (iterated Uehling)      & -0.00113 \\[1ex]
Self--Energy [Eq. (\ref{SEestimate})]
                                  & 0.00302 \\[1ex]
Salpeter correction [Eq.~(\ref{reslog})] 
                                  & 0.00038 \\[1ex]
\hline
Total                             & -1859.08986 \\[1ex]
\hline
\hline
\end{tabular}
\end{center}
\end{minipage}
\end{center}
\end{table}

%
%
\section{Numerical evaluation of QED corrections}

In order to provide a more complete picture of pionium we have
evaluated numerically a number of QED corrections to the 
1S level of pionium. We explicitly exclude QCD corrections 
whose evaluation represents a difficult separate 
problem~\cite{GaLyRuGa2001}.
We proceed as follows (see the sequence of the rows of Tab.~\ref{table1}):
\begin{itemize}
\item We start from the one-body Klein--Gordon energy for a
particle of (reduced) mass $m/2$, given in Eq.~(\ref{KGresult}),
including relativistic corrections of order $(Z\alpha)^4$.
\item We use a Klein-Gordon equation numerical
solver developed for pionic atoms which was developed
originally for the precise
evaluation of vacuum polarization corrections~\cite{LeBoGo1998,InSi2001}
in order to supplement the (almost neglible) difference 
between the exact one-body relativistic Klein--Gordon energy
and the $(Z\alpha)^4$-result from Eq.~(\ref{KGresult}),
thereby confirming the expression (\ref{KGresult})
for the relativistic correction.
\item The effect due to the pion Coulomb form factor is 
also included in an approximative framework
by replacing in the numerical solution of the Klein--Gordon
equation the Coulomb potential by the interaction potential of two
uniformly charged spheres of mean-square radius $R_{\mathrm{rms}}=0.61$~fm.
For the calculation,
we employ the radius and the pion mass from the particle
data group\cite{PDBook} (all other physical constants used in 
the evaluations come from the 1998
adjustment\cite{MoTa2000}).
\item We add as a ``two-body correction'' the difference
of the results from Eq.~(\ref{NLresult}) for the energy of the 
relativistic two-body system and the one-body result
given in Eq. (\ref{KGresult}).
\item The Uehling potential is evaluated in a relativistic
framework, using the relativistic
numerical equation solvers~\cite{LeBoGo1998,InSi2001}.
The result is in very good agreement with the nonrelativistic treatment
discussed in Sec.~\ref{VacuumEffects} (this is natural because $Z=1$).
\item The higher-order VP corrections attributed to
Wichmann--Kroll~\cite{WiKr1956} and
K{\"a}ll{\'e}n--Sabry~\cite{KaSa1955} are supplemented, as well as an
evaluation of the iterated loop-after-loop Uehling contribution to all
orders in $\alpha$.  The Wichmann--Kroll correction~\cite{WiKr1956} is
here negligible because $Z=1$.  VP potentials given in
Ref.~\cite{FuRi1976} and analytic expressions from \cite{Klars1977}
are used.  More details about the numerical procedure to evaluate
these corrections can be found in Ref.~\cite{BoIn2000}.
\item Finally, the effects due to the scalar
self-energy given in Eq.~(\ref{SEestimate}) and the 
relativistic recoil (Salpeter) correction listed in
Eq.~(\ref{reslog}) are added.
\end{itemize}
The main observation one can draw from Tab.~\ref{table1} is 
that the vacuum-polarization is 300 times larger than the
self-energy. Even the K{\"a}ll{\'e}n and Sabry correction, while of order
$\alpha^2$ is about twice as large as the self-energy correction. The
iterated Uehling correction, while dominated by terms of order
$\alpha^2$ is dominated in turn by the scalar self-energy.
Nevertheless, we stress here that the numbers contained in 
Tab.~\ref{table1} will be modified when the Klein--Gordon
equation is solved with the strong interaction potential 
incorporated directly into the equation solver. Analogously, 
one cannot avoid having to solve the Dirac equation exactly 
for high-$Z$ systems where the electron wave function rests
significantly inside the nucleus~\cite{MoSo1993}.
At present, in view of the formidable experimental
difficulties associated with a study of the atomic spectrum of 
pionium, we give the numbers in Tab.~\ref{table1} as an indication 
of the relative size and order-of-magnitude
of the specific QED corrections. 

\section{Conclusion}

We have presented in Sec.~\ref{BreitEffects}
a simplified derivation for the 
relativistic and recoil corrections of order $\alpha^4$
to a bound state of two spinless particles. The results
agree with previous calculations~\cite{Na1972}. As evident from 
equation~(\ref{hbreit}) and discussed in further 
detail in Sec.~\ref{BreitEffects}, the zitterbewegung term
is absent in a bound system of two spinless particles. 

The self-energy effect is suppressed in systems with spinless
particles in comparison to the vacuum polarization effect as
discussed in Secs.~\ref{VacuumEffects}
and~\ref{SelfEnergyEffects}, because 
the lightest known spinless particle is much heavier than the 
electron, which implies that the electronic vacuum 
polarization effect is larger 
by two orders of $\alpha$ than the self-energy effect in 
bound systems of spinless particles. We provide a
complete result for the leading scalar-QED correction
in pionium
of order $\alpha^5\,\ln\alpha$ in Sec.~\ref{SelfEnergyEffects}. 
A list of further QED corrections to the 1S level of pionium is
presented in Tab.~\ref{table1}.

%
%
\section*{Acknowledgments}

U.D.J.~acknowledges helpful and elucidating discussions with 
K. Pachucki, and very kind hospitality extended during 
an extended stay at the Laboratoire Kastler~Brossel, where part of this
work was completed. He would also like to acknowledge support
from the Deutscher Akademischer Austauschdienst (DAAD).
G.S.~acknowledges support from BMBF and GSI. Laboratoire 
Kastler~Brossel is Unit\'{e} Mixte de Recherche
associ\'{e}e au CNRS No.~8552.

\end{document}